\documentclass[envcountsame,envcountreset,envcountsect]{llncs}
\usepackage{amsmath}
\usepackage{amssymb}
\usepackage{ifthen}
\usepackage{url}

\newcommand{\abs}[1]{\left|\mathinner{#1}\right|}

\newcommand{\mediumSize}[1]{\fontsize{8pt}{10pt}\selectfont %mk
#1\normalsize} %mk%pw
\newcommand{\mediumFont}[1]{\normalfont\mediumSize{#1}}
\newcommand{\malcev}%
  {\mathop{\text{\normalsize{\raisebox{0.1mm}{\textcircled{\raisebox{0.2mm}{\mediumFont{\textit{m}}}}}}}}}%mk
\font\petite=cmmi10 at 8pt
\def\malcev{\mathbin{\hbox{$\bigcirc$\rlap{\kern-9pt\raise0,75pt\hbox{\petite m}}}}}

\newcommand{\Alpha}{\mathsf{alph}}

\newcommand{\varietyFont}[1]{\mathrm{\mathbf{#1}}}
\newcommand{\DA}{\varietyFont{D\hspace{-1pt}A}}
\newcommand{\LI}{\varietyFont{LI}}

\newcommand{\Jone}{\varietyFont{J_1}}

\let\J\Jtrivial
\newcommand{\RR}{\varietyFont{R}}
\newcommand{\LL}{\varietyFont{L}}
\newcommand{\X}{\varietyFont{X}}
\let\V\Vara
\let\W\Varb
\newcommand{\VarFO}{\varietyFont{FO}}
\newcommand{\VarTL}{\varietyFont{TL}}
\newcommand{\K}{\varietyFont{K}}
\newcommand{\D}{\varietyFont{D}}

\newcommand{\LangV}{\mathcal{V}}  %%% generic language variety

\newcommand{\LangFO}{\mathcal{FO}}
\newcommand{\LangTL}{\mathcal{TL}}
\newcommand{\underLangTL}{\underline{\mathcal{TL}}}
\def\calJ{\mathcal{J}}
\def\calL{\mathcal{L}}
\def\calR{\mathcal{R}}

\newcommand{\greenFont}[1]{\mathcal{#1}}
\newcommand{\gR}{\mathrel{\greenFont{R}}}
\newcommand{\gL}{\mathrel{\greenFont{L}}}
\newcommand{\gJ}{\mathrel{\greenFont{J}}}

\newcommand{\FO}{\mathsf{FO}} %{\mathrm{FO}}  %%% first order logic

\newcommand{\TL}{\mathsf{TL}}  %%% local temporal logic
\newcommand{\PTL}{\mathsf{PTL}}  %%% local temporal logic

\newcommand{\XX}{\mathrm{\mathsf{X}}}
\newcommand{\YY}{\mathrm{\mathsf{Y}}}
\newcommand{\ZZ}{\mathrm{\mathsf{Z}}}

\newcommand{\TRUE}{\mathbf{\top}}

\newcommand{\RIGHT}{\mathrel{\triangleright}}
\newcommand{\LEFT}{\mathrel{\triangleleft}}

\def\cqfd{\skip10=\parfillskip\parfillskip=0pt
\enspace\hfill\symbolecqfd\par\parfillskip=\skip10\par\medskip}
\def\symbolecqfd{\rlap{$\sqcap$}$\sqcup$}

\def\eop{\cqfd\endtrivlist}

\def\sketchproof{\rm \trivlist \item[\hskip \labelsep{\bf
Sketch of proof.}]}

\newcounter{commentcounter}

\def\inv{^{-1}}
\let\phi\varphi
\let\epsilon\varepsilon
\def\ord{\mathsf{ord}}

\def\underR{\underline{R}}
\def\underTL{\underline{\TL}}
\def\underVarTL{\underline{\VarTL}}
\def\llbracket{[\![}
\def\rrbracket{]\!]}

\begin{document}

\title{On $\FO^2$ quantifier alternation over words%
\thanks{Both authors acknowledge support from the ANR project
\textsc{dots}, the ESF program \textsc{AutoMathA} and the Indo-French
P2R project \textsc{modiste-cover}.}
}
\titlerunning{On $\FO^2$ quantifier alternation over words}  % abbreviated title (for running head)
\author{Manfred Kuf{}leitner\inst{1} \and Pascal 
Weil\inst{2,3}}
\authorrunning{M.~Kuf{}leitner and P.~Weil}   % abbreviated author list (for running head)
\tocauthor{Manfred Kuf{}leitner, Pascal Weil}
\institute{Institut f\"ur Formale Methoden der Informatik, 
Universit\"at Stuttgart, Germany
\and
LaBRI, Universit\'e de Bordeaux and CNRS, France
\and
Department of Computer Science and Engineering, IIT Delhi, India\\
\email{manfred.kufleitner@fmi.uni-stuttgart.de}\\
\email{pascal.weil@labri.fr}
}
\maketitle              % typeset the title of the contribution

\begin{abstract}
    We show that each level of the quantifier alternation hierarchy
    within $\FO^2[<]$ on words is a variety of languages.  We use the
    notion of condensed rankers, a refinement of the rankers defined
    by Weis and Immerman, to produce a decidable hierarchy of
    varieties which is interwoven with the quantifier alternation
    hierarchy -- and conjecturally equal to it.  It follows that the
    latter hierarchy is decidable within one unit, a much more precise
    result than what is known about the quantifier alternation
    hierarchy within $\FO[<]$, where no decidability result is known
    beyond the very first levels.
\end{abstract}

First-order logic is an important object of study in connection with
computer science and language theory, not least because many important
and natural problems are first-order definable: our understanding of
the expressive power of this logic and the efficiency of the solution
of related algorithmic problems are of direct interest in such fields
as verification.  Here, by first-order logic, we mean the first-order
logic of the linear order, $\FO[{<}]$, interpreted on finite words.

In this context, there has been continued interest in fragments of
first-order logic, defined by the limitation of certain resources,
e.g. the quantifier alternation hierarchy (which is closely related
with the dot-depth hierarchy of star-free languages).  It is still an
open problem whether each level of this hierarchy is
decidable.\footnote{On the other hand, the quantifier alternation
hierarchy collapses at level 2 for the first-order logic of the
successor $\FO[S]$ \cite{Thomas1982jcss,Pin2005dm}.}
Another natural restriction concerns the number of variables used (and
re-used!)  in a formula.  It is interesting, notably because the
trade-off between formula size and number of variables is known to be
related with the trade-off between parallel time and number of
processes, see
\cite{WeisImmerman2007csl,Immerman1999book,AdlerImmerman2003tcl,GroheSchweikardt2005lmcs}.

In this paper, we concentrate on $\FO^2[{<}]$, the 2-variable fragment
of $\FO[{<}]$.  It is well-known that every $\FO[{<}]$-formula is
logically equivalent with a formula using only 3 variables, but that
$\FO^2[{<}]$ is properly less expressive than $\FO[{<}]$.  The
expressive power of $\FO^2[{<}]$ was characterized in many interesting
fashions (see
\cite{SchwentickTV2001dlt,TessonTherien2002,TessonTherien2007lmcs,DiekertKufleitner2007dlt}),
and in particular, we know how to decide whether an $\FO[{<}]$-formula
is equivalent to one in $\FO^2[{<}]$.

A recent result of Weis and Immerman refined a result of Schwentick,
Th\'erien and Vollmer \cite{SchwentickTV2001dlt} to give a combinatorial
description of the $\FO^2_{m}[{<}]$-definable languages (those that
can be defined by an $\FO^2[{<}]$-formula with quantifier alternation
bounded above by $m$), using the notion of rankers. Rankers are 
finite sequences of instructions of the form \textit{go to the next 
$a$-position to the right} (resp. \textit{left}) \textit{of the current 
position}.

Our first set of results shows that $\LangFO^2_m$ (the
$\FO^2_m[{<}]$-definable languages), and the classes of languages
defined by rankers having $m$ alternations of directions (right
\textit{vs.} left), are varieties of languages.  This means that
membership of a language $L$ in these classes depends only on the
syntactic monoid of $L$, which justifies an algebraic approach of
decidability.

Our investigation shows that rankers are actually better suited to
characterize a natural hierarchy within unary temporal logic, and we
introduce the new notion of a condensed ranker, that is more adapted
to discuss the quantifier alternation hierarchy within $\FO^2[{<}]$.
There again, the alternation of directions in rankers defines
hierarchies of varieties of languages $\calR_m$ and $\calL_m$, with
particularly interesting properties.  Indeed, we show that these
varieties are decidable, that they admit a neat characterization in
terms of closure under deterministic and co-deterministic products,
and that $\calR_m \cup \calL_m \subseteq \LangFO^2_m \subseteq
\calR_{m+1}\cap\calL_{m+1}$.  The latter containments show that we can
effectively compute, given a language $L\in \LangFO^2$, an integer $m$
such that $L$ is in $\LangFO^2_{m+1}$, possibly in $\LangFO^2_{m}$,
but not in $\LangFO^2_{m-1}$.  This is much more precise than the
current level of knowledge on the general quantifier alternation
hierarchy in
$\FO[{<}]$.\footnote{%
Unfortunately, it does not help with the general problem since a
language $L$ is $\FO^2[{<}]$-definable if and only if $L$ and its
complement are $\Sigma_2$-definable \cite{PinWeil1997tocs}.}

%%%%%%%%%%%%
\section{An algebraic approach to study $\FO^2_m$}

If $u\in A^+$ is a non-empty word, we denote by $u[i]$ the letter of
$u$ in position $i$ ($1 \le i \le |u|$), and by $u[i,j]$ be the factor
$u[i]\cdots u[j]$ of $u$ ($1 \le i \le j \le |u|$).  Then we identify
the word $u$ with the logical structure $(\{1,\ldots, |u|\},
(\mathbf{a})_{a\in A})$, where $\mathbf{a}$ denotes the set of
integers $i$ such that $u[i] = a$.

Let $\FO[{<}]$ (resp.  $\FO^k[{<}]$, $k\ge 0$) denote the set of
first-order formulas using the unary predicates $\mathbf{a}$ ($a\in
A$) and the binary predicate $<$ (resp.  and at most $k$ variable
symbols).  It is well-known that $\FO^3[{<}]$ is as expressive as
$\FO[{<}]$ and that $\FO^2[{<}]$ is properly less expressive.

In the sequel, we omit specifying the predicate $<$ and we write
simply $\FO$ or $\FO^k$.  The classes of $\FO$- and $\FO^2$-definable
languages have well-known beautiful characterizations
\cite{SchwentickTV2001dlt,TessonTherien2002,TessonTherien2007lmcs,DiekertKufleitner2007dlt}.
Two are of particular interest in this paper.

- The algebraic characterization in terms of recognizing monoids: a
language is $\FO$-definable if and only if it is recognized by a
finite aperiodic monoid, i.e., one in which $x^n = x^{n+1}$ for each
element $x$ and for all $n$ large enough (Sch\"utzenberger and
McNaughton-Ladner, see \cite{Straubing1994book}); and a language is
$\FO^2$-definable if and only if it is recognized by a finite monoid
in $\DA$ (see \cite{TessonTherien2002}), a class of monoids with many
interesting characterizations, which will be discussed later.  These
algebraic characterizations prove the decidability of the
corresponding classes of languages: $L$ is $\FO$ (resp.  $\FO^2$)
definable if and only if the (effectively computable) syntactic monoid
of $L$ is in the (decidable) class of aperiodic monoids (resp.  in
$\DA$).

- The language-theoretic characterization: a language is in
$\FO$-definable if and only if it is star-free, i.e., it can be
obtained from singletons using Boolean operations and concatenation
products (Sch\"utzenberger, see \cite{Pin1986book}); a language is
$\FO^2$-definable if and only if it can be written as the disjoint
union of unambiguous products of the form $B_0^*a_1B_1^*\cdots
a_kB_k^*$, where $k\ge 0$, the $a_i$ are letters and the $B_i$ are
subsets of the alphabet.  Such a product is called
\textit{unambiguous} if each word $u \in B_0^*a_1B_1^*\cdots a_kB_k^*$
admits a unique factorization in the form $u = u_0a_1u_1\cdots a_ku_k$
such that $u_i\in B_i^*$ for each $i$.

We now concentrate on $\FO^2$-formulas and we define two important
parameters concerning such formulas.  To simplify matters, we consider
only formulas where negation is used only on atomic formulas so that,
in particular, no quantifier is negated.  This is naturally possible
up to logical equivalence.  Now, with each formula $\phi\in \FO^2$,
we associate in the natural way a parsing tree: each occurrence of a
quantification, $\exists x$ or $\forall x$, yields a unary node, each
occurrence of $\lor$ or $\land$ yields a binary node, and the leaves
are labeled with atomic or negated atomic formulas.  Each path from
root to leaf in this parsing tree has a \textit{quantifier label},
which is the sequence of quantifier node labels ($\exists$ or
$\forall$) encountered along this path.  A \textit{block} in this
quantifier label is a maximal factor consisting only of $\exists$ or
only of $\forall$.  The \textit{quantifier depth} of $\phi$ is the
maximum length of the quantifier label of a path in the parsing tree
of $\phi$, and the \textit{number of blocks} of $\phi$ is the maximum
number of blocks in the quantifier label of a path in its parsing
tree.

We let $\FO^2_{m,n}$ denote the set of first-order formulas with
quantifier depth at most $n$ and with at most $m$ blocks and let
$\FO^2_m$ denote the union of the $\FO^2_{m,n}$ for all $n$.  We also
denote by $\LangFO^2$ ($\LangFO^2_m$) the class of $\FO^2$
($\FO^2_m$)-definable languages.
Weis and Immerman's characterization of the expressive power of
$\FO^2_{m,n}[{<}]$ \cite{WeisImmerman2007csl} in terms of rankers, see
Theorem~\ref{thm: IW refined} below, forms the basis of our own results.

%%%%%%%%%%%%
\subsection{Rankers and logic}

A \textit{ranker} \cite{WeisImmerman2007csl} is a non-empty word on
the alphabet $\{\XX_a,\YY_a \mid a\in A\}$.\footnote{%
Weis and Immerman write $\RIGHT_a$ and $\LEFT_a$ instead of $\XX_a$
and $\YY_a$.  We rather follow the notation in
\cite{DiekertKufleitner2007dlt}, where $\XX$ and $\YY$ refer to the
future and past operators of \textsf{LTL}.}
Rankers may define positions in words: given a word $u \in A^+$ and a
letter $a\in A$, we denote by $\XX_a(u)$ (resp.  $\YY_a(u)$) the least
(resp.  greatest) integer $1\le i \le |u|$ such that $u[i] = a$.  If
$a$ does not occur in $u$, we say that $\YY_a(u)$ and $\XX_a(u)$ are
not defined.  If in addition $q$ is an integer such that $1 \le q \le
|u|$, we let
\begin{align*}
    \XX_a(u,q) &= \XX_a(u[q+1,|u|]) \\
    \YY_a(u,q) &= \YY_a(u[1,q-1]).
\end{align*}
These definitions are extended to all rankers: if $r'$ is a ranker,
$\ZZ \in \{\XX_a,\YY_a \mid a\in A\}$ and $r = r'\ZZ $, we let $r(u,q)
= \ZZ(u,r'(u,q))$ if $r'(u,q)$ and $\ZZ(u,r'(u,q))$ are defined, and
we say that $r(u,q)$ is undefined otherwise.

Finally, if $r$ starts with an $\XX$- (resp.  $\YY$-) letter, we say
that $r$ defines the position $r(u) = r(u,0)$ (resp.  $r(u) =
r(u,|u|+1)$), or that it is undefined on $u$ if this position does not
exist.  Then $L(r)$ is the language of all words on which $r$ is
defined.  We say that the words $u$ and $v$ \textit{agree on a class}
$R$ of rankers if exactly the same rankers from $R$ are defined on $u$
and $v$.

The \textit{depth} of a ranker $r$ is defined to be its length (as a
word).  A \textit{block} in $r$ is a maximal factor in $\{\XX_a \mid
a\in A\}^+$ (an $\XX$-block) or in $\{\YY_a \mid a\in A\}^+$ (a
$\YY$-block).  If $n\ge m$, we denote by $R^\XX_{m,n}$ (resp.
$R^\YY_{m,n}$ ) the set of $m$-block, depth $n$ rankers, starting with
an $\XX$ -(resp.  $\YY$-) block, and we let $R_{m,n} = R^\XX_{m,n}
\cup R^\YY_{m,n}$ and $\underR^\XX_{m,n} = \bigcup_{n'\le
n}R^\XX_{m,n'} \cup \bigcup_{m'<m,n'<n}R_{m',n'}$.  We define
$\underR^\YY_{m,n}$ dually and we let $\underR^\XX_m = \bigcup_{n\ge
m}\underR^\XX_{m,n}$, $\underR^\YY_m = \bigcup_{n\ge
m}\underR^\YY_{m,n}$ and $\underR_m = \underR^\XX_m \cup
\underR^\YY_m$.

%%%%%%%%%%%%%%%%%%
\subsubsection{Rankers and temporal logic}
Let us depart for a moment from the consideration of $\FO^2$-formulas,
to observe that rankers are naturally suited to describe the different
levels of a natural class of temporal logic.  The symbols $\XX_a$ and
$\YY_a$ ($a\in A$) can be seen as modal (temporal) operators, with the
\textit{future} and \textit{past} semantics respectively.  We denote
the resulting temporal logic (known as \textit{unary temporal logic})
by $\TL$: its only atomic formula is $\TRUE$, the other formulas are
built using Boolean connectives and modal operators.  Let $u\in A^+$
and let $0 \le i \le |u|+1$.  We say that $\TRUE$ holds at every
position $i$, $(u,i) \models \TRUE$; Boolean connectives are
interpreted as usual; and $(u,i) \models \XX_a\phi$ (resp.
$\YY_a\phi$) if and only if $(u,j) \models \phi$, where $j$ is the
least $a$-position such that $i < j$ (resp.  the greatest $a$-position
such that $j < i$).  We also say that $u \models \XX_a\phi$ (resp.
$\YY_a\phi$) if $(u,0) \models \XX_a\phi$ (resp.  $(u,1+|u|) \models
\YY_a\phi$).

$\TL$ is a fragment of \textit{propositional temporal logic} $\PTL$;
the latter is expressively equivalent to $\FO$ and $\TL$ is
expressively equivalent to $\FO^2$, see \cite{TessonTherien2002}.

As in the case of $\FO^2$-formulas, one may consider the parsing tree
of a $\TL$-formula and define inductively its depth and number of
alternations (between past and future operators).  If $n\ge m$, the
fragment $\TL^\XX_{m,n}$ (resp.  $\TL^\YY_{m,n}$) consists of the
$\TL$-formulas with depth $n$ and with $m$ alternations, in which
every branch (of the parsing tree) with exactly $m$ alternations
starts with future (resp.  past) operators.  The fragments
$\TL_{m,n}$, $\underTL^\XX_{m,n}$, $\underTL^\YY_{m,n}$,
$\underTL^\XX_{m}$, $\underTL^\YY_{m}$ and $\underTL_m$ are defined
according to the same pattern as in the definition of $R_{m,n}$,
$\underR^\XX_{m,n}$, $\underR^\YY_{m,n}$, $\underR^\XX_{m}$,
$\underR^\YY_{m}$ and $\underR_m$.  We also denote by
$\LangTL^\XX_{m,n}$ ($\LangTL^\XX_m$, $\underLangTL_m$, etc) the class
of $\TL^\XX_{m,n}$ ($\TL^\XX_m$, $\underLangTL^\XX_m$, etc)-definable
languages.  The following result is elementary.

\begin{proposition}\label{prop: temporal}
    Let $1 \le m \le n$.  Two words satisfy the same $\TL^\XX_{m,n}$
    formulas if and only if they agree on rankers from $R^\XX_{m,n}$.
    A language is in $\LangTL^\XX_{m,n}$ if and only if it is a 
    Boolean combination of languages of the form $L(r)$, $r\in 
    R^\XX_{m,n}$.
    
    Similar statements hold for $\TL^\YY_{m,n}$, $\TL_{m,n}$,
    $\underTL^\XX_{m,n}$, $\underTL^\YY_{m,n}$, $\underTL^\XX_{m}$,
    $\underTL^\YY_{m}$ and $\underTL_m$, relative to the corresponding
    classes of rankers.
\end{proposition}

%%%%%%%%%%%%%%%%%%
\subsubsection{Rankers and $\FO^2$}
The connection established by Weis and Immerman
\cite{WeisImmerman2007csl} between rankers and formulas in
$\FO^2_{m,n}$, Theorem~\ref{thm: IW refined} below, is deeper.  If
$x,y$ are integers, we let $\ord(x,y)$, the \textit{order type} of $x$
and $y$, be one of the symbols $<$, $>$ or $=$, depending on whether
$x<y$, $x>y$ or $x=y$.

\begin{theorem}\label{thm: IW refined}
    Let $u,v \in A^*$ and let $1\le m \le n$. Then $u$ and $v$ satisfy the 
    same formulas in $\FO^2_{m,n}$ if and only if
    \begin{itemize}
	\item[\textbf{(WI}] \textbf{1)}\enspace $u$ and $v$ agree on
	rankers from $R_{m,n}$,
	
	\item[\textbf{(WI}] \textbf{2)}\enspace if the rankers $r \in
	\underR_{m,n}$ and $r' \in \underR_{m-1,n-1}$ are defined on
	$u$ and $v$, then $\ord(r(u),r'(u)) = \ord(r(v),r'(v))$.
	
	\item[\textbf{(WI}] \textbf{3)}\enspace if $r \in
	\underR_{m,n}$ and $r' \in \underR_{m,n-1}$ are defined on $u$
	and $v$ and end with different direction letters, then
	$\ord(r(u),r'(u)) = \ord(r(v),r'(v))$.
    \end{itemize}
\end{theorem}  

\begin{corollary}\label{TLm in FO2m}
    For each $n\ge m\ge 1$, $\underLangTL_{m,n} \subseteq
    \LangFO^2_{m,n}$ and $\underLangTL_m \subseteq \LangFO^2_m$.
\end{corollary}

%%%%%%%%%%%%%%%%%%
\subsubsection{$\FO^2_m$ and $\underTL_m$-definable languages form varieties}
Our first result is the following.  We refer the reader to
\cite{Pin1986book} and to Section~\ref{sec: varieties} below for
background and discussion on varieties of languages.

\begin{proposition}\label{prop: logical varieties}
    For each $n\ge m\ge 1$, the classes $\underLangTL^\XX_{m,n}$
    $\underLangTL^\YY_{m,n}$, $\underLangTL^\YY_m$,
    $\underLangTL^\YY_m$, $\underLangTL_{m,n}$,
    $\underLangTL_m$, $\LangFO^2_{m,n}$ and $\LangFO^2_m$ are
    varieties of languages.
\end{proposition}

\sketchproof
Let $\rho_{m,n}$ be the relation for two words to agree on
$\underTL^\XX_{m,n}$-formulas.  Using Proposition~\ref{prop:
temporal}, one verifies that $\rho_{m,n}$ is a finite index
congruence.  Then a language is $\underTL^\XX_{m,n}$-definable if and
only if it is a union of $\rho_{m,n}$-classes, if and only if it is
recognized by the finite monoid $A^*/\!\rho_{m,n}$.  It follows that
these languages are exactly those accepted by the monoids in the
pseudovariety generated by the $A^*/\!\rho_{m,n}$, for all finite
alphabets $A$, and hence they form a variety of languages.

The proof for the other fragments of $\TL$ is similar.  For the
fragments of $\FO^2$, we use Theorem~\ref{thm: IW refined} instead of
Proposition~\ref{prop: temporal}.
\eop

This result shows that, for a given regular language $L$,
$\underTL^\XX_m$- (resp.  $\underTL_m$-, $\FO^2_m$-, etc) definability
is characterized algebraically, that is, it depends only on the
syntactic monoid of $L$.  This justifies using the algebraic path to
tackle decidability of these definability problems.  Eilenberg's
theory of varieties provides the mathematical framework.

%%%%%%%%%%%%%%%%%%
\subsection{A short survey on varieties and pseudovarieties}\label{sec: varieties}

We summarize in this section the information on monoid and variety
theory that will be relevant for our purpose, see
\cite{Pin1986book,Almeida1994book,TessonTherien2002,TessonTherien2007lmcs}
for more details.

A language $L\subseteq A^*$ is \textit{recognized} by a monoid $M$ if
there exists a morphism $\phi\colon A^* \rightarrow M$ such that $L =
\phi\inv(\phi(L))$.  For instance, if $u\in A^*$ and $B\subseteq A$,
let $\Alpha(u) = \{a\in A \mid u = vaw \textrm{ for some $v,w\in
A^*$}\}$ and $[B] = \{u\in A^* \mid \Alpha(u) = B\}$.  Then $[B]$ is
recognized by the direct product of $|B|$ copies of the 2-element
monoid $\{0,1\}$ (multiplicative).

A \textit{pseudovariety} of monoids is a class of finite monoids
closed under taking direct products, homomorphic images and
submonoids.  Pseudovarieties of subsemigroups are defined similarly.
A \textit{class of languages} $\LangV$ is a collection $\LangV =
(\LangV(A))_A$, indexed by all finite alphabets $A$, such that
$\LangV(A)$ is a set of languages in $A^*$.  If $\V$ is a
pseudovariety of monoids, we let $\LangV(A)$ be the set of languages
of $A^*$ recognized by a monoid in $\V$.  The class $\LangV$ is closed
under Boolean operations, residuals and inverse homomorphic images.
Classes of recognizable languages with these properties are called
\textit{varieties} of languages, and Eilenberg's theorem (see
\cite{Pin1986book}) states that the correspondence $\V \mapsto \LangV$,
from pseudovarieties of monoids to varieties of languages, is
one-to-one and onto.  Moreover, the decidability of membership in the
pseudovariety $\V$, implies the decidability of the variety $\LangV$:
indeed, a language is in $\LangV$ if and only if its (effectively
computable) syntactic monoid is in $\V$.

For every finite semigroup $S$ and $s \in S$, we denote by $s^\omega$
the unique power of $s$ which is idempotent.  The \textit{Green
relations} are another important concept to describe monoids: if $S$
is a monoid and $s,t\in S$, we say that $s\le_{\gJ}t$ (resp.  $s
\le_{\gR} t$, $s \le_{\gL} t$) if $s = utv$ (resp.  $s = tv$, $s =
ut$) for some $u,v \in S$.  We also say that $s \gJ t$ is $s\le_{\gJ}
t$ and $t\le_{\gJ} s$.  The relations $\gR$ and $\gL$ are defined
similarly.

Pseudovarieties that will be important in this paper are the 
following.

- $\Jone$, the pseudovariety of idempotent and commutative monoids,
whose corresponding variety of languages consists of the Boolean
combinations of languages of the form $[B]$.

- $\RR$, $\LL$ and $\J$, the pseudovarieties of $\gR$-, $\gL$- and 
$\gJ$-trivial monoids; a monoid is, say, $\gR$-trivial if each of its 
$\gR$-classes is a singleton.

- $\DA$, the pseudovariety of all monoids in which $(xy)^\omega x
(xy)^\omega = (xy)^\omega$ for all $x,y$;  $\DA$ has a great many
characterizations in combinatorial, algebraic and logical terms
\cite{Almeida1994book,PinWeil1997tocs,SchwentickTV2001dlt,TessonTherien2002,TessonTherien2007lmcs}.
% - $\DA$ is the pseudovariety of all monoids in which $(xy)^\omega x
% (xy)^\omega = (xy)^\omega$ for all $x,y$.  $\DA$ has a great many
% characterizations in combinatorial, algebraic and logical terms
% \cite{Almeida1994book,PinWeil1996ja,SchwentickTV2001dlt,TessonTherien2002,TessonTherien2007lmcs}.

- $\K$ (resp. $\D$, $\LI$) is the pseudovariety of semigroups in 
which $x^\omega y = x^\omega$ (resp. $yx^\omega = x^\omega$, 
$x^\omega yx^\omega = x^\omega$) for all $x,y$.

Finally, if $\V$ is a pseudovariety of semigroups and $\W$ is a
pseudovariety of monoids, we say that a finite monoid $M$ lies in the
\textit{Mal'cev product} $\W\malcev\V$ if there exists a finite monoid
$T$ and onto morphisms $\alpha\colon T\rightarrow M$ and $\beta\colon
T\rightarrow N$ such that $N\in \W$ and $\beta\inv(e) \in \V$ for each
idempotent $e$ of $N$.  Then $\W\malcev\V$ is a pseudovariety of
monoids and we have in particular \cite{Pin1986book,Almeida1994book,PinWeil1996ja}:
$$\K\malcev\Jone = \K\malcev\J = \RR,\enspace \D\malcev\Jone =
\D\malcev\J = \LL, \enspace\LI\malcev\Jone = \LI\malcev\J = \DA.$$
We denote by $\underVarTL^\XX_{m,n}$ $\underVarTL^\YY_{m,n}$,
$\underVarTL^\YY_m$, $\underVarTL^\YY_m$, $\underVarTL_{m,n}$,
$\underVarTL_m$, $\VarFO^2_{m,n}$ and $\VarFO^2_m$ the pseudovarieties
corresponding to the language varieties discovered in
Proposition~\ref{prop: logical varieties}.

%%%%%%%%%%%%%%%%%%
\section{Main results}\label{sec: main}

Our main tool to approach the decidability of $\FO^2_m$-definability
lies in a variant of rankers, which we borrow from a proof in Weis and
Immerman's paper \cite{WeisImmerman2007csl}.  As in the turtle
language of \cite{SchwentickTV2001dlt}, a ranker can be seen as a
sequence of instructions: go to the next $a$ to the right, go to the
next $b$ to the left, etc.  We say that a ranker $r$ is
\emph{condensed on $u$} if it is defined on $u$, and if the sequence
of positions visited \textit{zooms in} on $r(u)$, never crossing over
a position already visited.
Formally, $r = \ZZ_{1} \cdots \ZZ_{n}$ is condensed on $u$ if there
exists a chain of open intervals
$$(0,\abs{u}+1) = (i_0, j_0) \supset (i_1,j_1) \supset \cdots \supset
(i_{n-1},j_{n-1}) \ni r(u)$$
such that for all $1 \leq \ell \leq n-1$ the following properties are
satisfied:
\begin{itemize}
    \item If $\ZZ_{\ell} \ZZ_{\ell + 1} = \XX_a \XX_b$ then
    $(i_\ell,j_\ell) = (\XX_a(u,i_{\ell-1}), j_{\ell-1})$.
    \item If $\ZZ_{\ell} \ZZ_{\ell + 1} = \YY_a \YY_b$ then
    $(i_\ell,j_\ell) = (i_{\ell-1}, \YY_a(u,j_{\ell-1})$.
    \item If $\ZZ_{\ell} \ZZ_{\ell+1} = \XX_a \YY_b$ then
    $(i_\ell,j_\ell) = (i_{\ell-1}, \XX_a(u,i_{\ell-1}))$.
    \item If $\ZZ_{\ell} \ZZ_{\ell+1} = \YY_a \XX_b$ then
    $(i_\ell,j_\ell) = (\YY_a(u,j_{\ell-1}), j_{\ell-1})$.
\end{itemize}

For instance, the ranker $\XX_a\YY_b\XX_c$ is defined on the words
$bac$ and $bca$, but it is condensed only on $bca$.  Rankers in
$\underR_1$, or of the form $\XX_a\YY_{b_1}\cdots\YY_{b_k}$ or
$\YY_a\XX_{b_1}\cdots\XX_{b_k}$, are condensed on all words on which
they are defined.  We denote by $L_c(r)$ the set of all words on which
$r$ is condensed.

Condensed rankers form a natural notion, which is equally well-suited
to the task of describing $\FO^2_m$-definability (see
Theorem~\ref{thm: IW refined 2} below).  With respect to $\TL$, for
which Proposition~\ref{prop: temporal} shows a perfect match with the
notion of rankers, they can be interpreted as adding a strong notion
of unambiguity, see Section~\ref{sec: unambiguous} below and the work
of Lodaya, Pandya and Shah \cite{LodayaPS2008ifip} on unambiguous
interval temporal logic.

%%%%%%%%%%%%%%%%%%
\subsection{Condensed rankers determine a hierarchy of pseudovarieties}

Let us say that two words $u$ and $v$ \textit{agree on condensed
rankers from a set} $R$ of rankers, if the same rankers are condensed
on $u$ and $v$.  We write $u \RIGHT_{m,n} v$ (resp.  $u \LEFT_{m,n}
v$) if $u$ and $v$ agree on condensed rankers in $\underR^\XX_{m,n}$
(resp.  $\underR^\YY_{m,n}$).

These relations turn out to have a very nice recursive
characterization.  For each word $u\in A^*$ and letter $a$ occurring
in $u$, the \textit{$a$-left} (resp.  \textit{$a$-right})
\textit{factorization} of $u$ is the factorization that isolates the
leftmost (resp.  rightmost) occurrence of $a$ in $u$; that is, the
factorization $u = u_-au_+$ such that $a$ does not occur in $u_-$
(resp.  $u_+$).  We say that the word $a_1\cdots a_r$ is a
\textit{subword} of $u$ if $u$ can be factored as $u = u_0a_1u_1\cdots
a_ru_r$, with the $u_i\in A^*$.

\begin{proposition}\label{prop: ppties congruences}
    The relations $\RIGHT_{m,n}$ and $\LEFT_{m,n}$ ($n \ge m \ge 1$)
    are uniquely determined by the following properties.
    
    - $u \RIGHT_{1,n} v$ if and only if $u \LEFT_{1,n} v$, if and only
    if $u$ and $v$ have the same subwords of length at most $n$.
    
    - If $m \ge 1$, then $u \RIGHT_{m,n} v$ if and only if $\Alpha(u) 
    = \Alpha(v)$, $u \LEFT_{m-1,n-1} v$ and for each letter $a\in 
    \Alpha(u)$, the $a$-left factorizations $u = u_-au_+$ and $v = 
    v_-av_+$ satisfy $u_- \LEFT_{m-1,n-1} v_-$ and $u_+  
    \RIGHT_{m,n-1} v_+$.

    - If $m \ge 1$, then $u \LEFT_{m,n} v$ if and only if $\Alpha(u) 
    = \Alpha(v)$, $u \RIGHT_{m-1,n-1} v$ and for each letter $a\in 
    \Alpha(u)$, the $a$-right factorizations $u = u_-au_+$ and $v = 
    v_-av_+$ satisfy $u_+ \RIGHT_{m-1,n-1} v_+$ and $u_- 
    \LEFT_{m,n-1} v_-$.
\end{proposition}

% This can be used to prove the following.

\begin{corollary}\label{right and left congruences}
    The relations $\RIGHT_{m,n}$ and $\LEFT_{m,n}$ are finite-index 
    congruences.
\end{corollary}

For each $m \ge 1$, let us denote by $\RR_m$ (resp.  $\LL_m$) the
pseudovariety generated by the quotients $A^*/\!\RIGHT_{m,n}$ (resp.
$A^*/\!\LEFT_{m,n}$), where $n\ge m$ and $A$ is a finite alphabet.
Corollary~\ref{right and left congruences} shows that a language $L$
is in the corresponding variety $\calR_m$ (resp.  $\calL_m$) if and
only if $L$ is a Boolean combination of languages of the form
$L_c(r)$, with $r\in \underR^\XX_m$ (resp.  $\underR^\YY_m$).

By definition, for all $m\ge 1$, $\RR_m$ and $\LL_m$ are contained in
both $\RR_{m+1}$ and $\LL_{m+1}$.  According to the first statement of
Proposition~\ref{prop: ppties congruences}, $\RIGHT_{1,n} =
\LEFT_{1,n}$ is the congruence defining the piecewise $n$-testable
languages studied by Simon in the early 1970s, and that, in
consequence, $\RR_1 = \LL_1 = \J$, the pseudovariety of
$\calJ$-trivial monoids \cite{Pin1986book}.

In addition, one can show that if a position in a word $u$ is defined by a
ranker $r \in \underR^\XX_{m,n}$ (resp.  $\underR^\YY_{m,n}$), then
the same position is defined by a ranker $s \in \underR^\XX_{m,n}$
(resp.  $\underR^\YY_{m,n}$) which is condensed on $u$.  This leads to
the following result.

\begin{proposition}\label{prop: TL vs Rm 1}
    Let $n \ge m \ge 1$.  If the words $u$ and $v$ agree on condensed
    rankers in $\underR^\XX_{m,n}$ (resp.  $\underR^\YY_{m,n}$), then
    they agree on rankers from the same class.  In particular,
    $\underVarTL^\XX_m \subseteq \RR_m$ and $\underVarTL^\YY_m
    \subseteq \LL_m$
\end{proposition}

As indicated above, condensed rankers allow for a description of
$\FO^2_m$-defin\-ability, as neat as with ordinary rankers: more
precisely, we show that the statement of Weis and Immerman's theorem
can be modified to used condensed rankers instead.

\begin{theorem}\label{thm: IW refined 2}
    Let $u,v \in A^*$ and let $1\le m \le n$. Then $u$ and $v$ satisfy the 
    same formulas in $\FO^2_{m,n}$ if and only if
    \begin{itemize}
	\item[\textbf{(WI}] \textbf{1c)}\enspace $u$ and $v$ agree on
	condensed rankers from $R_{m,n}$,
	
	\item[\textbf{(WI}] \textbf{2c)}\enspace if the rankers $r \in
	\underR_{m,n}$ and $r' \in \underR_{m-1,n-1}$ are condensed on
	$u$ and $v$, then $\ord(r(u),r'(u)) = \ord(r(v),r'(v))$.
	
	\item[\textbf{(WI}] \textbf{3c)}\enspace if $r \in
	\underR_{m,n}$ and $r' \in \underR_{m,n-1}$ are condensed on
	$u$ and $v$ and end with different direction letters, then
	$\ord(r(u),r'(u)) = \ord(r(v),r'(v))$.
    \end{itemize}
\end{theorem}  

Thus there is a connection between $\LangFO^2_m$ and the varieties
$\calR_m$ and $\calL_m$.  But much more can be said about the latter
varieties.

%%%%%%%%%%%%%%%%%%
\subsection{Language hierarchies}\label{sec: language hierarchies}

Proposition~\ref{prop: ppties congruences} also leads to a
description of the language varieties $\calR_m$ and $\calL_m$ in terms
of deterministic and co-deterministic products.  Recall that a product
of languages $L = L_0a_1L_1\cdots a_kL_k$ ($k \ge 1$, $a_i \in A$,
$L_i \subseteq A^*$) is said to be \textit{deterministic} if, for $0
\le i \le k$, each word $u\in L$ has a unique prefix in
$L_0a_1L_1\cdots L_{i-1}a_i$. If for each $i$, the letter $a_i$ does 
not occur in $L_{i-1}$, the product $L_0a_1L_1\cdots a_kL_k$ is 
called \textit{visibly deterministic}: this is obviously a particular 
case of a deterministic product.

The definition of a \textit{co-deterministic} or \textit{visibly
co-deterministic} product is dual, in terms of suffixes instead of
prefixes.  If $\LangV$ is a class of languages and $A$ is a finite
alphabet, let $\LangV^{det}(A)$ (resp.  $\LangV^{vdet}(A)$,
$\LangV^{codet}(A)$, $\LangV^{vcodet}(A)$) be the set of all Boolean
combinations of languages of $\LangV(A)$ and of deterministic (resp.
visibly deterministic, co-deterministic, visibly co-deterministic)
products of languages of $\LangV(A)$.  Sch\"utzenberger gave algebraic
characterizations of the closure operations $\LangV \longmapsto
\LangV^{det}$ and $\LangV \longmapsto \LangV^{codet}$, see
\cite{Pin1986book}: if $\LangV$ is a variety of languages and if $\V$
is the corresponding pseudovariety of monoids, then $\LangV^{det}$ and
$\LangV^{codet}$ are varieties of languages and the corresponding
pseudovarieties are, respectively, $\K\malcev\V$ and $\D\malcev\V$. 
Then we show the following.

\begin{proposition}\label{malcev statement}
    For each $m\ge 1$, we have $\calR_{m+1} = \calL_m^{vdet} =
    \calL_m^{det}$, $\RR_{m+1} = \K \malcev \LL_m$, $\calL_{m+1} =
    \calR_m^{vcodet} = \calR_m^{codet}$ and $\LL_{m+1} = \D \malcev
    \RR_m$.  In particular, $\RR_2 = \RR$ and $\LL_2 = \LL$.
\end{proposition}

\sketchproof
Proposition~\ref{prop: ppties congruences} shows that $\calR_{m+1}
\subseteq \calL_m^{vdet}$, which is trivially contained in
$\calL_m^{det}$. The last containment is proved algebraically, by 
showing that if $\gamma\colon A^* \to M$ is an onto morphism, and $M 
\in \K \malcev \LL_m$, then for some large enough $n$, $u 
\RIGHT_{m+1,n} v$ implies $\gamma(u) = \gamma(v)$: thus $M$ is a 
quotient of $A^*/\!\RIGHT_{m+1,n}$ and hence, $M \in \RR_{m+1}$. This 
proof relies on a technical property of semigroups in $\DA$: if $a\in 
A$ occurs in $\Alpha(v)$ and $\gamma(u) \gR \gamma(uv)$, then 
$\gamma(uva) \ gR \gamma(u)$.
\eop

It turns out that the $\RR_m$ and the $\LL_m$ were studied in the
semigroup-theoretic literature (Kufleitner, Trotter and Weil,
\cite{TrotterWeil1997au,KufleitnerWeil2009submitted}).  In
\cite{KufleitnerWeil2009submitted}, it is defined as the hierarchy of
pseudovarieties obtained from $\J$ by repeated applications of the
operations $\X\mapsto \K\malcev\X$ and $\X \mapsto \D\malcev\X$.
Proposition~\ref{malcev statement} shows that it is the same hierarchy
as that considered in this paper%
\footnote{More precisely, the pseudovarieties $\RR_m$ and $\LL_m$ in
\cite{KufleitnerWeil2009submitted} are pseudovarieties of semigroups,
and the $\RR_m$ and $\LL_m$ considered in this paper are the classes
of monoids in these pseudovarieties.}%
.  The following results are proved in
\cite[Section 4]{KufleitnerWeil2009submitted}.

\begin{proposition}\label{prop from sf}
    The hierarchies $(\RR_m)_m$ and $(\LL_m)_m$ are infinite chains of
    decidable pseudovarieties, and their unions are equal to $\DA$.
    Moreover, every $m$-generated monoid in $\DA$ lies in
    $\RR_{m+1}\cap\LL_{m+1}$.
\end{proposition}

The decidability statement in Proposition~\ref{prop from sf} is in
fact a consequence of a more precise statement (see
\cite{TrotterWeil1997au,KufleitnerWeil2009submitted}) which gives
defining pseudoidentities for the $\RR_m$ and $\LL_m$.  Let
$x_1,x_2,\dots$ be a sequence of variables.  If $u$ is a word on that
alphabet, we let $\bar u$ be the mirror image of $u$, that is, the
word obtained from reading $u$ from right to left.  We let
\begin{align*}
    &G_2 = x_2x_1,\qquad I_2 = x_2x_1x_2,\\
    \hbox{for $n > 2$,}\quad &G_n = x_n\overline{G_{n-1}},\qquad I_n =
    G_nx_n\overline{I_{n-1}},\\
    &\phi(x_1) = (x_1^\omega x_2^\omega x_1^\omega)^\omega,\quad \phi(x_2) =
    x_2^\omega,\\
    \hbox{ and, for $n > 2$, }\quad&\phi(x_n) = (x_n^\omega
    \phi(\overline{G_{n-1}}G_{n-1})^\omega x_n^\omega)^\omega.
\end{align*}
Then we have \cite{KufleitnerWeil2009submitted}:

\begin{proposition}
    For each $m\ge 2$, $\RR_m = \DA \cap \llbracket \phi(G_m) =
    \phi(I_m)\rrbracket$ and $\LL_m = \DA \cap \llbracket
    \phi(\overline{G_m}) = \phi(\overline{I_m})\rrbracket$.
\end{proposition}

\begin{example}
    For $\RR_2$, this yields the pseudo-identity
    $x_2^\omega(x_1^\omega x_2^\omega x_1^\omega)^\omega =
    x_2^\omega(x_1^\omega x_2^\omega x_1^\omega)^\omega x_2^\omega$.
    One can verify that, together with the pseudo-identity defining 
    $\DA$, this is equivalent to the usual pseudo-identity
    describing $\RR = \RR_2$, namely $(st)^\omega s = (st)^\omega$.
    
    For $\RR_3 = \K\malcev\LL$, no pseudo-identity was known in the
    literature.  We get
    \begin{align*}
	\phi(G_3) &= (x_3^\omega ((x_1^\omega x_2^\omega
	x_1^\omega)^\omega x_2^\omega (x_1^\omega x_2^\omega
	x_1^\omega)^\omega)^\omega x_3^\omega)^\omega \\
	\phi(I_3) &= (x_3^\omega ((x_1^\omega x_2^\omega x_1^\omega)^\omega
	x_2^\omega (x_1^\omega x_2^\omega x_1^\omega)^\omega)^\omega
	x_3^\omega)^\omega \\
	& \qquad \qquad (x_3^\omega ((x_1^\omega x_2^\omega
	x_1^\omega)^\omega x_2^\omega (x_1^\omega x_2^\omega
	x_1^\omega)^\omega)^\omega x_3^\omega)^\omega x_2^\omega
	(x_1^\omega x_2^\omega x_1^\omega)^\omega x_2^\omega.
    \end{align*}
% %     
%     $$G_3 = x_3x_1x_2,\ I_3 = x_3x_1x_2x_3x_2x_1x_2$$
% %     
%     $$\phi(x_3) = (x_3^\omega \phi(x_1x_2x_2x_1)^\omega
%     x_3^\omega)^\omega$$
% %     
%     $$\phi(G_3) = (x_3^\omega ((x_1^\omega x_2^\omega
%     x_1^\omega)^\omega x_2^\omega (x_1^\omega x_2^\omega
%     x_1^\omega)^\omega)^\omega x_3^\omega)^\omega$$
% %     
%     $$\phi(I_3) = (x_3^\omega ((x_1^\omega x_2^\omega
%     x_1^\omega)^\omega x_2^\omega (x_1^\omega x_2^\omega
%     x_1^\omega)^\omega)^\omega x_3^\omega)^\omega (x_3^\omega
%     \phi(x_1x_2x_2x_1)^\omega x_3^\omega)^\omega x_2^\omega
%     (x_1^\omega x_2^\omega x_1^\omega)^\omega x_2^\omega.$$
% 
\end{example}

%%%%%%%%%%%%%%%%%%
\subsection{Connection with the $\VarTL_m$ and the $\VarFO^2_m$  hierarchies}

Proposition~\ref{prop: TL vs Rm 1} established a containment between
the $\RR_m$ (resp.  $\LL_m$) and the $\underVarTL_m$ hierarchies.  A
technical analysis allows us to prove a containment in the other
direction, but one that is not very tight -- showing the difference
between the consideration of condensed rankers and that of ordinary
rankers.

\begin{proposition}\label{prop 3/2}
    $\RR_2 = \underVarTL^\XX_2$ and $\LL_2 \subseteq
    \underVarTL^\YY_2$.
    If $m \ge 3$ and if two words agree on rankers in
    $\underR^\XX_{\lfloor3m/2\rfloor}$ (resp.
    $\underR^\YY_{\lfloor3m/2\rfloor}$), then they agree on condensed
    rankers in $\underR^\XX_m$ (resp.  $\underR^\YY_m$).  In
    particular $\RR_m \subseteq \underVarTL^\XX_{\lfloor 3m/2\rfloor}$
    and $\LL_m \subseteq \underVarTL^\YY_{\lfloor 3m/2\rfloor}$.
\end{proposition}

\begin{example}\label{TL3 vs R3}
    The language $L_c(\XX_a\YY_b\XX_c)$ is in $\calR_3$ 
    and not in $\underLangTL_3^\XX$.
\end{example}
% 
% 
% % %%%%%%%%%%%%%%%%%%
% % \subsection{Interwoven hierarchies}
% 
The connection between the $\RR_m$, $\LL_m$ and $\VarFO^2_m$
hierarchies is tighter.

\begin{theorem}\label{thm: interwoven}
    Let $m\ge 1$. Every language in $\calR_m$ or $\calL_m$ is 
    $\FO^2_m$-definable, and every $\FO^2_m$-definable language is in 
    $\calR_{m+1} \cap \calL_{m+1}$. Equivalently, we have
    $$\RR_m \vee \LL_m \subseteq \VarFO^2_m \subseteq \RR_{m+1} \cap 
    \LL_{m+1},$$
    where $\V\vee\W$ denotes the least pseudovariety containing $\V$ 
    and $\W$.
\end{theorem}

\sketchproof
The containment $\RR_m \vee \LL_m \subseteq \VarFO^2_m$ follows
directly from Property (\textbf{WI 1c}) in Theorem~\ref{thm: IW
refined 2}.  The proof of the converse containment also relies on that
theorem.  We show that if $u \RIGHT_{m+1,2n}$ or $u \LEFT_{m+1,2n}$,
then Properties (\textbf{WI 1c}), (\textbf{WI 2c}) and (\textbf{WI
3c}) hold for $m,n$.  This is done by a complex and quite technical
induction.
\eop

If $m=1$, we know that $\RR_2 \cap \LL_2 = \RR \cap \LL = \J =
\RR_1\vee\LL_1$: this reflects the elementary observation that
$\FO^2_1$-definable languages, like $\FO_1$-definable languages, are
the piecewise testable languages.  For $m\ge 2$, we conjecture that
$\RR_m \vee \LL_m$ is properly contained in $\RR_{m+1} \cap
\LL_{m+1}$.  The following shows it holds for $m = 2$.

\begin{example}\label{example R2L2 vs FO2}
    $L = \{b,c\}^*ca\{a,b\}^*$ is $\FO^2_2$-definable, by the
    following formula:
    \begin{eqnarray*}
	\exists i && (\mathbf{c}(i) \land (\forall j\ (j<i \to 
	\neg\mathbf{a}(j))) \land (\forall j\ (j>i \to 
	\neg\mathbf{c}(j)))) \\
	\land\enspace \exists i && (\mathbf{a}(i) \land (\forall j\ (j<i \to 
	\neg\mathbf{a}(j))) \land (\forall j\ (j>i \to 
	\neg\mathbf{c}(j))))\\
	\land\enspace \forall i && (\mathbf{b}(i) \to (\exists j\ (j<i \land 
	\mathbf{a}(j)) \lor (\exists j\ (j>i \land \mathbf{c}(j)))).
    \end{eqnarray*}
    The words $u_n = (bc)^n(ab)^n$ are in $L$, while the words $v_n =
    (bc)^nb(ca)^n$ are not.  Almeida and Azevedo showed that
    $\RR_2\vee\LL_2$ is defined by the pseudo-identity
    $(bc)^\omega(ab)^\omega = (bc)^\omega b(ab)^\omega$ \cite[Theorem
    9.2.13 and Exercise 9.2.15]{Almeida1994book}).  In particular, for
    each language $K$ recognized by a monoid in $\RR_2\vee\LL_2$, the
    words $u_n$ and $v_n$ (for $n$ large enough) are all in $K$, or
    all in the complement of $K$.  Therefore $L$ is not recognized by
    such a monoid, which proves that $\RR_2\vee\LL_2$ is strictly 
    contained in $\VarFO^2_2$, and hence also in $\RR_3\cap\LL_3$. It 
    also shows that $\underLangTL_2$ is properly contained in 
    $\LangFO^2_2$.
\end{example}

Finally, we formulate the following conjecture.

\begin{conjecture}
    For each $m\ge 1$, $\VarFO^2_m = \RR_{m+1} \cap \LL_{m+1}$.
\end{conjecture}

%%%%%%%%%%%%%%%%%%
\section{Consequences}\label{sec: unambiguous}

% %%%%%%%%%%%%%%%%%%
% \subsection{Almost deciding quantifier alternation in $\FO^2$}

The main consequence we draw of Theorem~\ref{thm: interwoven} and of
the decidability of the pseudovarieties $\RR_m$ and $\LL_m$ is 
summarized in the next statement.

\begin{theorem}
    Given an $\FO^2$-definable language $L$, one can compute an
    integer $m$ such that $L$ is $\FO^2_{m+1}$-definable but not
    $\FO^2_{m-1}$-definable.
    That is: we can decide the quantifier alternation level of $L$
    within one unit.
\end{theorem}

\sketchproof
If $M \in \DA$, we can compute the largest $m$ such that $M \not\in
\RR_m \cap \LL_m$ (Proposition~\ref{prop from sf}).  Then $M \not\in
\VarFO^2_{m+1} \setminus \VarFO^2_{m-1}$ by Theorem~\ref{thm:
interwoven}.
\eop

% %%%%%%%%%%%%%%%%%%
% \subsection{Infinite and collapsing hierarchies}

The fact that the $\RR_m$ and $\LL_m$ form strict hierarchies
(Proposition~\ref{prop from sf}), together with Theorem~\ref{thm:
interwoven}, proves that the $\LangFO^2_m$ hierarchy is infinite.
Weis and Immerman had already proved this result by combinatorial
means \cite{WeisImmerman2007csl}, whereas our proof is algebraic.
From that result on the $\LangFO^2_m$, it is also possible to recover
the strict hierarchy result on the $\RR_m$ and $\LL_m$ and the fact
that their union is equal to $\DA$.

By the same token, Propositions~\ref{prop: TL vs Rm 1} and~\ref{prop 3/2}
show that the $\underLangTL_m$ (resp.  $\underVarTL_m$) hierarchy is
infinite and that its union is all of $\LangFO^2$ (resp.  $\DA$).

Similarly, the fact that an $m$-generated element of $\DA$ lies in
$\RR_{m+1}\cap\LL_{m+1}$ (Proposition~\ref{prop from sf}), shows that
an $\FO^2$-definable language in $A^*$ lies in
$\calR_{|A|+1}\cap\calL_{|A|+1}$, and hence in $\LangFO^2_{m+1}$ -- a
fact that was already established by combinatorial means by Weis and
Immerman \cite[Theorem 4.6]{WeisImmerman2007csl}.  It also shows that
such a language is in $\underLangTL_{\frac32(|A|+1)}$ by
Proposition~\ref{prop 3/2}.

% %%%%%%%%%%%%%%%%%%
% \subsection{Infinite hierarchies and unambiguous polynomials}\label{sec: unambiguous}

Finally we note the following refinement on \cite[Proposition
4.6]{KufleitnerWeil2009submitted}.  It was mentioned in the
introduction that the languages in $\LangFO^2$ are disjoint unions of 
unambiguous products of the form
$B_0^*a_1B_1^*\cdots a_kB_k^*$, where each $B_i$ is a subset of $A$.
Propositions~\ref{malcev statement} and~\ref{prop from sf} 
imply the following statement.

\begin{proposition}
    The least variety of languages containing the languages of the 
    form $B^*$ ($B \subseteq A$) and closed under visibly 
    deterministic and visibly co-deterministic products, is 
    $\LangFO^2$.
    
    Every unambiguous product of languages of the form
    $B_0^*a_1B_1^*\cdots a_kB_k^*$ (with each $B_i \subseteq A$), can
    be expressed in terms of the $B_i^*$ and the $a_i$ using only
    Boolean operations and at most $|A|+1$ applications of visibly
    deterministic and visibly co-deterministic products, starting with
    a visibly deterministic (resp.  co-deterministic) product.
\end{proposition}

The weaker statement with the word \textit{visibly} deleted was proved
by the authors in \cite{KufleitnerWeil2009submitted}, as well as by 
Lodaya, Pandya and Shah \cite{LodayaPS2008ifip}.

\bibliographystyle{plain}
{\small

\newcommand{\Ju}{Ju}\newcommand{\Th}{Th}\newcommand{\Yu}{Yu}

}

\end{document}